\documentclass[10pt,prb,twocolumn,showpacs,superscriptaddress]{revtex4-1}
\usepackage{graphicx}
\begin{document}

%\preprint{}

\title{Interplay between superconductivity and antiferromagnetism in some iron-pnictides single crystals studied by $^{57}$Fe M\"ossbauer spectroscopy\/}
     \author{J. Munevar}
     \author{H. Micklitz}
     \author{J. Ag\"uero}
     \affiliation{Centro Brasileiro de Pesquisas Fisicas, Rua Xavier Sigaud 150, Rio de Janeiro, Brazil}
     \author{C. L. Zhang}
     \affiliation{Department of Physics and Astronomy, The University of Tennessee, Knoxville, Tennessee 37996-1200, USA}
     \author{H. Q. Luo}
     \affiliation{Beijing National Laboratory for Condensed Matter Physics, Institute of Physics, Chinese Academy of Sciences, Beijing 100190, China}
     \author{C. Arg\"uello}
     \author{Y. J. Uemura}
     \affiliation{Department of Physics, Columbia University, New York, New York 10027, USA}
     \author{Pengcheng Dai}
     \affiliation{Department of Physics and Astronomy, The University of Tennessee, Knoxville, Tennessee 37996-1200, USA}
     \affiliation{Beijing National Laboratory for Condensed Matter Physics, Institute of Physics, Chinese Academy of Sciences, Beijing 100190, China}
     \author{E. Baggio-Saitovitch}
     \email[author to whom correspondences should be addressed: E-mail ]{elisa@cbpf.br}
     \affiliation{Centro Brasileiro de Pesquisas Fisicas, Rua Xavier Sigaud 150, Rio de Janeiro, Brazil}
     \date{\today}
   \begin{abstract}
\noindent
{We have performed detailed $^{57}$Fe M\"ossbauer spectroscopy measurements on Ba$_{0.78}$K$_{0.22}$Fe$_2$As$_2$ and BaFe$_{2-x}$Ni$_x$As$_2$ single crystal mosaics showing antiferromagnetic ordering below $T_N$ with superconductivity below $T_C$. Analysis of the M\"ossbauer spectra shows a decrease in the magnetic hyperfine (hf) field but no change in the magnetic volume fraction below $T_C$. This clearly indicates the coexistence of magnetism and superconductivity in these compounds. The decrease in the magnetic hf field below $T_C$ depends on the difference between $T_N$ and $T_C$, being the largest for $T_N$ close to $T_C$. Two different explanations for this observation are given. We also find that the non-magnetic volume fraction below $T_N$ correlates with the Ni doping $x$, being large for high $T_C$ and small for high $T_N$.  }
\end{abstract}
%\keywords{}
\maketitle

The coexistence of superconductivity (SC) and antiferromagnetism (AF) in the recently discovered iron-based pnictides is a heavily discussed subject. Since it is the general opinion that both phenomena originate from Fe-3d electrons, a competition between these two phenomena may be expected. Indeed, a decrease in the intensity of magnetic Bragg peaks below the superconducting transition temperature T$_C$ in BaFe$_{2-x}$(Co,Ni)$_x$As$_2$  has been observed \cite{pratt,christianson,mwang1,mwang2}. It has been interpreted as a reduction of the static AF-ordered Fe magnetic moments below T$_C$. Such a decrease clearly indicates that the static AF order is a competing phase to superconductivity. However, a decrease in the local magnetic field has not been observed in muon spin rotation ($\mu$SR) experiments on BaFe$_{2-x}$(Co,Ni)$_x$As$_2$ \cite{tomoprivate}, i.e. in experiments using a local probe, but a decrease has been recently observed in Ba$_{1-x}$K$_x$Fe$_2$As$_2$ polycrystalline samples through the same technique \cite{wiesenmayer}.  It is proposed that this result implies $s^{+-}$ pairing of the Cooper pairs, meaning unconventional superconductivity and coexistence of SDW antiferromagnetism and superconductivity\cite{rafael}.  The length scale of coexistence of the two phenomena, SC and AF order, may be the decisive parameter which can explain both results: phase separation on a mesoscopic length scale larger than the SC coherence length ( about 2 nm \cite{putti} ), for example, has been proposed for  Ba$_{1-x}$K$_x$Fe$_2$As$_2$ \cite{park}. If this also is the case for BaFe$_{2-x}$(Co,Ni)$_x$As$_2$, a change in the magnetic volume fraction at T$_C$, rather than a change in the static Fe magnetic moment, could explain both the neutron scattering as well as the $\mu$SR results for this compound. In a very recent paper, however, the coexistence of the two phenomena, SC and AF order, in Ba$_{1-x}$K$_x$Fe$_2$As$_2$ on a lattice parameter length scale has been claimed \cite{1104.0453}. The length scale of coexistence of the two phenomena, SC and AF order, in the above given iron-pnictides, therefore, is still an unsolved problem. With the hope to contribute to this problem we have performed detailed $^{57}$Fe M\"ossbauer studies above and below T$_C$ on single crystals of Ba$_{0.78}$K$_{0.22}$Fe$_2$As$_2$ as well as BaFe$_{2-x}$Ni$_x$As$_2$ in order to see if a change in the static Fe moment at T$_C$ can be reflected in local magnetic hyperfine field at the $^{57}$Fe nuclei in these compounds.   

Single crystal mosaics of Ba$_{0.78}$K$_{0.22}$Fe$_2$As$_2$ and Ba$_2$Fe$_{2-x}$Ni$_x$As$_2$ (x = 0.065, 0.075, 0.085) were used for M\"ossbauer studies.  Single crystal thin platelets were taken to mount circled mosaics with roughly 1 cm of diameter.  The details of crystal growth procedures are published elsewhere \cite{ychen}.  Mosaics were mounted with c-axis perpendicular to the absorber plane and parallel to 14.4 keV $\gamma$-rays from $^{57}$Co source.  M\"ossbauer spectra were taken in a variable temperature helium cryostat, allowing temperatures between 2 and 300 K. Both, M\"ossbauer source ($^{57}$Co:Rh), moving in a sinusoidal mode, and absorber have been kept at the same temperature. Isomer shifts are reported relative to that of $\alpha$-Fe.  %Separate single crystals were used to measure magnetic properties in a SQUID magnetometer with field perpendicular to c-axis, in order to confirm the superconducting response.  Those measurements are shown in the figure \ref{fig0}.%M\"ossbauer spectra are shown in the figure \ref{fig1}.

\begin{figure}[!t]
\includegraphics[width=8cm]{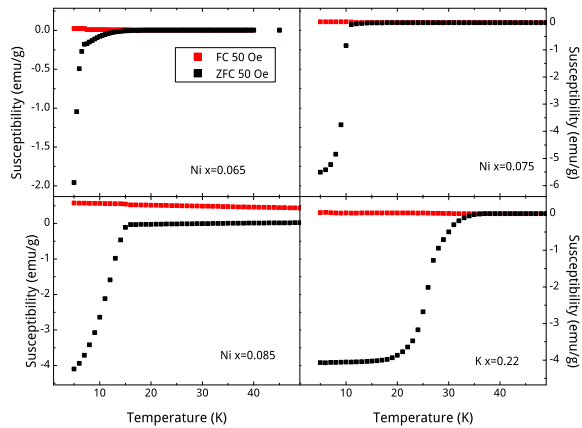}
\caption{\label{fig0} Magnetic susceptibility measurements performed on BaFe$_{2-x}$Ni$_x$As$_2$ and Ba$_{0.87}$K$_{0.22}$Fe$_2$As$_2$ single crystals, with 50 Oe applied field, parallel to $c$ axis.}
\end{figure}

Magnetic response of single crystals of BaFe$_{2-x}$Ni$_x$As$_2$ and Ba$_{0.78}$K$_{0.22}$Fe$_2$As$_2$ was measured in a SQUID magnetometer with an applied field of 50 Oe and field parallel to $c$ axis, taking 5 - 50 K as temperature range.  From these measurements (see fig \ref{fig0}) a diamagnetic response related to Meissner effect due to the onset of superconducting ordering is clearly observed.  The transition temperatures are around 7 K for Ni doped 0.065 sample, 12 K for Ni doped 0.075 sample, 15 K for Ni doped 0.085 sample and 33 K for K doped single crystal.  The measurement for BaFe$_{1.935}$Ni$_{0.065}$As$_2$ crystal shows a slow decrease of the signal around 15 K and a sharp and large diamagnetic response below 7 K; we believe the sharp decrease corresponds to the bulk of the sample, while the other response we observe corresponds to the sample surface, modified due to contact with air.

\begin{figure}[!t]
\includegraphics[width=8cm]{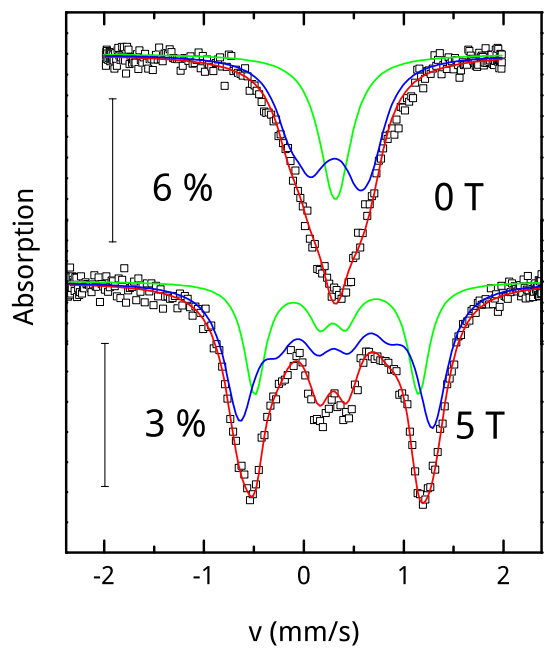}
\caption{\label{fig1} $^{57}$Fe M\"ossbauer spectra for BaFe$_{1.925}$Ni$_{0.075}$As$_2$ single crystal mosaic taken at 4.35 K.  An external magnetic field of 5 T was applied, parallel to crystal c-axis.}
\end{figure}

The choice of the model used for M\"ossbauer spectra analysis was extracted from the spectra shown in figure \ref{fig1} for BaFe$_{1.925}$Ni$_{0.075}$As$_2$ at 4.35 K under applied external field (5 T), as compared with the spectrum without external field (0 T).  When no external field is applied, the spectrum could be fitted with an incommensurate spin density wave distribution \cite{1106.1332},  or as a superposition of a magnetic and nonmagnetic component (66 \% - 33 \% area proportions), or even taking a single component with a large linewidth value due to inequivalent lattice sites caused by Ni doping.  All those approximations can reproduce the measured spectra.  When external field of 5 T is applied parallel to c-axis, the spectrum show two sites: one subspectrum with $B_{hf}$ = 5.1(2) T, $\theta$ = 0 degrees and 33 \% of relative area, and another subspectrum with $B_{hf}$ = 6.0(2) T, $\theta$ = 34(4) degrees and 66 \% of relative area.  The first component is exactly the nonmagnetic fraction of the sample under the external magnetic field, while the second component is found to be a superposition between the intrinsic Fe moment oriented parallel to the ab-plane and the external applied field.  Isomer shift and quadrupole splitting are essentially the same in both cases, with $\delta=$ 0.41(1) mm/s and $\Delta E_Q=$ 0.05(4) mm/s for both sites.  Fits with only one site can also work for both spectra, but large linewidth values are difficult to be satisfactorily explained within the model.  In a two site model, the nonmagnetic regions can be related to cancellation of magnetic moments, short range magnetism, or doping effects and distortion of the SDW magnetic structure.  Consideration of a magnetic and a nonmagnetic site comes naturally from the spectra fits performed, making our model simple and reliable.  It has been shown that probably the best model is assuming a field distribution related to an incommensurate spin density wave \cite{1106.1332}, taking a Fourier series expansion for the hyperfine field term.  In our case, a two site model will let us extract information on the interplay of SC and magnetism in these compounds, which is the goal of this work, and give physically reasonable information on our samples.  However, we should remark that using the simple two site model, the nonmagnetic fraction we obtain from such a fit, will give us an upper limit for the nonmagnetic fraction.  In the SDW model, part of the magnetic fraction will have very low magnetic field values and, therefore, the nonmagnetic fraction in the SDW model will be somewhat smaller.

\begin{figure*}[!t]
\includegraphics[width=16cm]{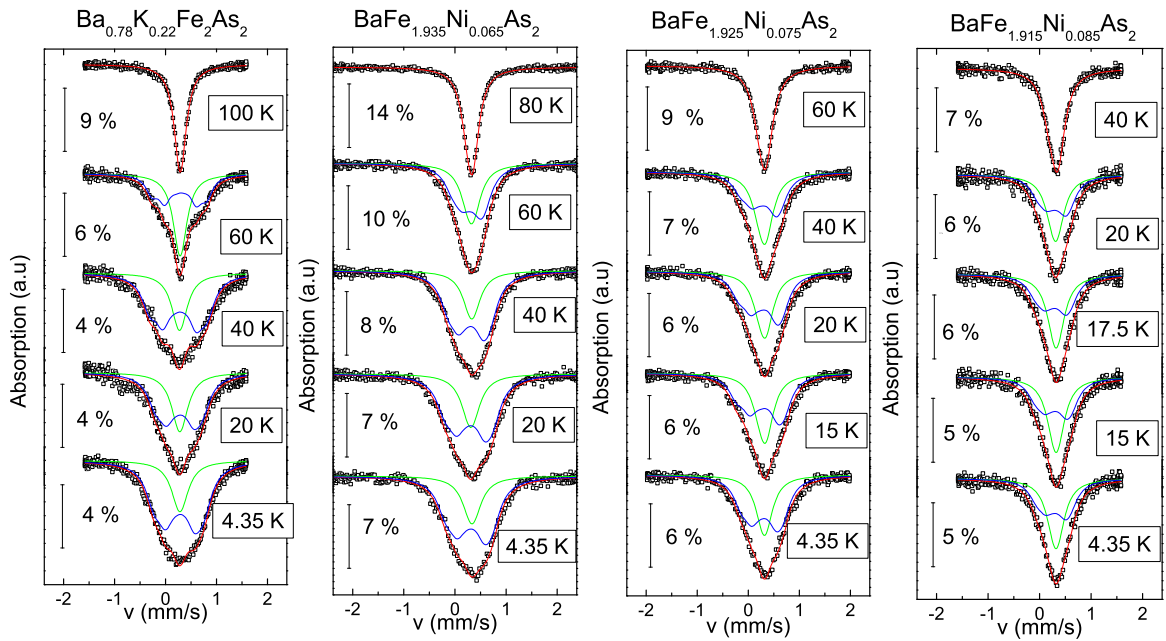}
\caption{\label{fig2} $^{57}$Fe M\"ossbauer spectra for Ba$_{0.78}$K$_{0.22}$Fe$_2$As$_2$ and BaFe$_{2-x}$Ni$_x$As$_2$ (x=0.065,0.075,0.085) single crystal mosaics.  $\gamma$ ray is parallel to crystal c-axis.  Fits assuming a two site model are shown, with a magnetic (blue lines) and a nonmagnetic (green single line) component below T$_N$.}
\end{figure*}

Above $T_N$, $^{57}$Fe M\"ossbauer spectra shown in figure \ref{fig2}, exhibit a single line which had been fitted with an unresolved quadrupole doublet. The $^{57}$Fe M\"ossbauer spectra taken below T$_N$ have been fitted with a non-magnetic and a magnetic components, having the same isomer shift and quadrupole interaction. The non-magnetic component was considered with an unresolved quadrupole doublet similar to the one of paramagnetic state, while the magnetic component was fitted within the full Hamiltonian model \cite{fhamilt}, essentially having seven fitting parameters such as magnetic hyperfine (hf) field, quadrupole interaction, angle between magnetic hf field and $\gamma$-ray direction (parallel to c-axis) as well as that between main component V$_{zz}$ of electric field gradient tensor and $\gamma$-ray direction, isomer shift, linewidth and line intensity. Three of these parameters have been kept fixed: quadrupole interaction below $T_N$ (will be discussed in detail later), angle between V$_{zz}$ and $\gamma$-ray direction, and linewidth; for the latter one we have taken the values obtained from the paramagnetic spectrum fit. The angle between V$_{zz}$ and the $\gamma$-ray direction was found to be close to zero in the paramagnetic state, as it is expected for V$_{zz}$ being parallel to the c-axis. This may induce an asymmetry in the absorption lines, which is also taken into account in the fitting process for single crystal measurements.

From the fits described above, we can extract the isomer shift for both sites, the magnetic hyperfine field and the nonmagnetic volume fraction.  Isomer shift for all samples studied are found to be close to $\delta=$0.39(1) mm/s.  A minor change in its value can be followed through Ni doping, since for x=0.065 we have $\delta=$ 0.40(1) mm/s and for x=0.085 we have $\delta=$ 0.38(1) mm/s.  Quadrupole splittings are well known for their capability to reflect structural phase transitions affecting the Fe electric interactions.  The measurements with applied field for Ni doped 0.075 sample and the spectra for Ni doped 0.065 sample show a change in the quadrupole interaction below $T_N$, related to the structural transition observed for this family of compounds \cite{mariella}.  A reduction of quadrupole splitting to 50 \% of the value above $T_N$ is observed.  For the other Ni doped samples we had to fix this quadrupole interaction, because a free fit would give unphysical results.  Its value was fixed to 50 \% of the value above $T_N$ assuming the behavior for this family of samples being similar in all cases.  For the K-doped sample we did not find any indication of change in the quadrupole interaction, even knowing that a structural transition indeed exists. This should indicate that K doping induces less disorder in FeAs tetrahedra rather than Co doping at Fe site, and this may influence all the properties of the crystals studied.  Linewidths never exceeded 0.4 mm/s.  From isomer shift values we assume Fe to be in +2 valence state, with small variations expected by adding holes/electrons to Fe valence band.  The angle between Fe hyperfine field and $z$ component of the electric field gradient of the lattice $V_{ZZ}$, called $\theta$, was fixed to 90 degrees, accordingly to SDW antiferromagnetic ordering with Fe moments lying in the $a,b$ plane, present in these samples.

\begin{figure}[!t]
\includegraphics[width=9cm]{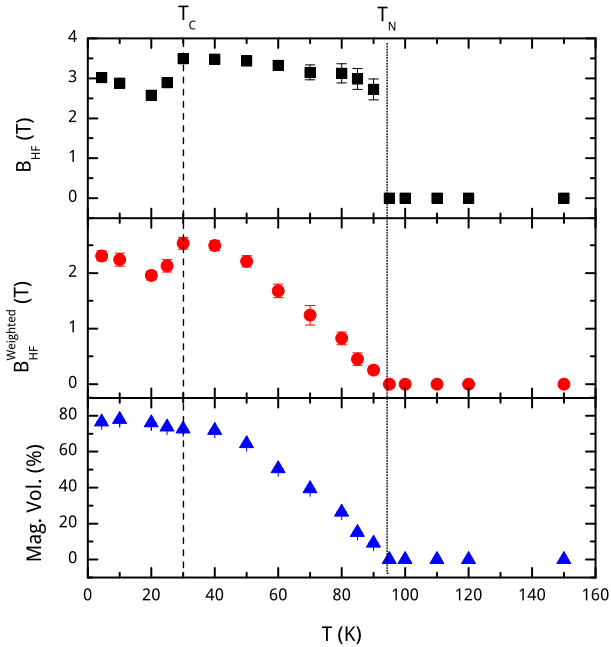}
\caption{\label{fig3} Magnetic hyperfine field $B_{hf}$, weighted $B_{hf}$ and nonmagnetic volume fraction extracted from $^{57}$Fe M\"ossbauer spectra fits for Ba$_{0.78}$K$_{0.22}$Fe$_2$As$_2$ single crystal mosaics.}
\end{figure}

Magnetic hyperfine field $B_{hf}(T)$, magnetic volume fraction and weighted magnetic hyperfine (hf) field, defined as the product of the magnetic hyperfine field and the magnetic volume fraction, are shown for Ba$_{0.78}$K$_{0.22}$Fe$_2$As$_2$ in figure \ref{fig3}.  Magnetic ordering is observed around 95 K showing a sharp increase of the magnetic moment for Fe.  Nevertheless, the magnetic fraction of the sample starts to increase slowly below $T_N$, reaching a steady value only below 40 K.  Below $T_C$ (~30 K) a decrease in $B_{hf}$ is observed.  The magnetic moment estimated from $B_{hf}$ for Fe\cite{dubiel} is around 0.2 $\mu_B$, indication of itinerant magnetism for Fe due to Fermi surface nesting \cite{rafael,chan,fawang}.  If we compare this value with that reported for BaFe$_2$As$_2$ ($\mu_{Fe}=0.36\mu_B$ \cite{bonville}) we have to conclude that the decrease in the magnetic moment is being induced by K doping on Ba site, that is, by distortion of the Fermi surface resulting in a reduction of the nesting between hole and electron pockets.  The weighted magnetic hf field shows a temperature dependence of the Fe moment which looks like that observed in neutron diffraction studies of a ordinary second order phase transition.  With a microscopic method (M\"ossbauer) we can distinguish between size of magnetic moment and magnetic volume fraction. With a macroscopic method (magnetization, neutron scattering) one can measure only the product of these 2 quantities. For that reason the magnetic phase transition looks like an ordinary second order phase transition if one uses a macroscopic method, while it is a ``local'' first order transition if one uses a microscopic method.  It is important to notice that nonmagnetic volume fraction does not reach 100 \% of the volume.  This can be caused either by local inhomogeneities due to K doping leading to several Fe sites with minor differences but strongly affecting the Fe coupling, or a slow crossover from short range to long range ordering caused by doping and distorting the electronic surfaces responsible for magnetism \cite{drew}.

\begin{figure}[!t]
\includegraphics[width=9cm]{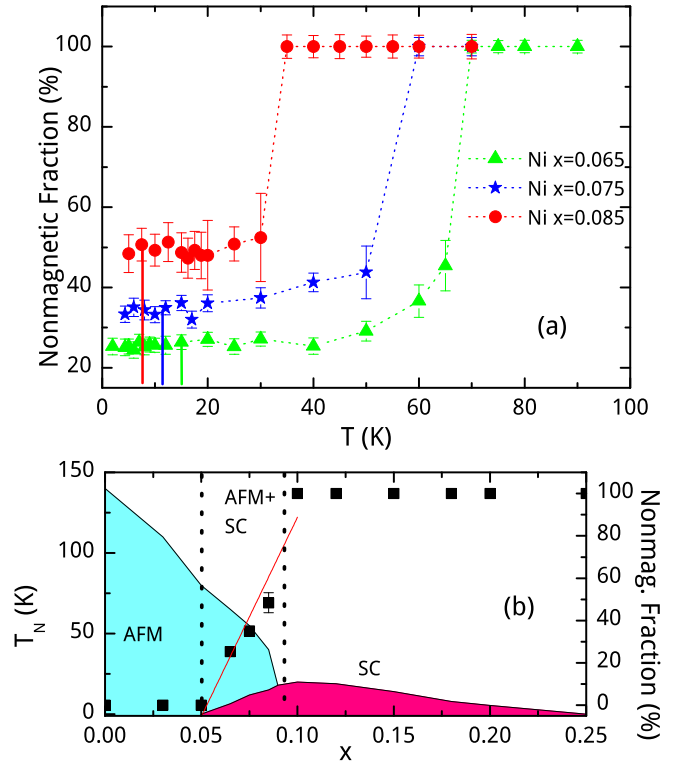}
\caption{\label{fig4} (a) Nonmagnetic volume fractions obtained from $^{57}$Fe M\"ossbauer spectra for BaFe$_{2-x}$Ni$_{x}$As$_2$ single crystal mosaics.  Lines drawn from data to bottom axis are indicating $T_C$ for each sample.  (b) Nonmagnetic volume fraction as a function of doping x.  The phase diagram is shown for comparison purposes.  The red straight line through the data is only a guide to the eyes. The region between the dotted lines in the phase diagram shows coexistence of magnetism and superconductivity, while the rest of the diagram show either magnetic or superconducting regions. }
\end{figure}

Figure \ref{fig4} shows the nonmagnetic volume fractions extracted from the M\"ossbauer fits for BaFe$_{2-x}$Fe$_x$As$_2$ single crystals mosaics.  In this figure it is clearly seen that just below $T_N$ an almost constant value of the nonmagnetic volume fraction is obtained, contrary to what is observed for the Ba$_{0.78}$K$_{0.22}$Ni$_2$as$_2$ single crystals (see fig. \ref{fig3}).  We do not observe any variation of the nonmagnetic volume fraction at or below $T_C$ in any of our studied samples.  This, however, one should expect if one wants to have a consistent explanation for both $\mu$SR and neutron scattering data (see above).  We also can observe from figure \ref{fig4}(a) that there exists a almost linear relation between the nonmagnetic volume fraction obtained from our fits and Ni doping (x) in the region where magnetism and superconductivity coexists in these samples (see figure \ref{fig4}(b)) and confirm that the appearance of the nonmagnetic fraction matches with the region where magnetism and superconducitivity coexist in the phase diagram of BaFe$_{2-x}$Fe$_x$As$_2$ single crystals\cite{mwang1}.  For optimal doped samples we obtain no magnetic response and for the underdoped crystals below the superconducting regime we expect a full magnetic order of the sample.

%From the magnetic hyperfine field results shown in the figure \ref{fig3}, we notice that a similar behavior had been seen in the same Ni doped single crystals through neutron scattering \cite{pratt,christianson,mwang1,mwang2}, in this work is reported a spin resonance and a decrease of the Bragg peaks intensity at $T_C$, meaning correlations between superconducting and magnetic ordering coming from the same electrons, this is, coexistence of magnetism and superconductivity.  The variation we observe in our magnetic hyperfine field data at $T_C$ seems to be directly related to the difference between $T_C$ and $T_N$, the shorter the difference between $T_C$ and $T_N$ the larger the variation of the magnetic hyperfine field.  A possible reason is that the energy scale for magnetic and superconducting ordering may tend to become of the same order as $T_C$ approaches $T_N$, the competition between SC and AF order becomes stronger.  This may also have a relation to the paramagnetic volume fraction (figure \ref{fig4}) at lower temperatures for different doping, because the same electrons may be responsible for magnetism and superconductivity in the system. %, and despite we find regions where no magnetic ordering is present, a correlation still persists.

\begin{figure}[t!h]
\includegraphics[width=9cm]{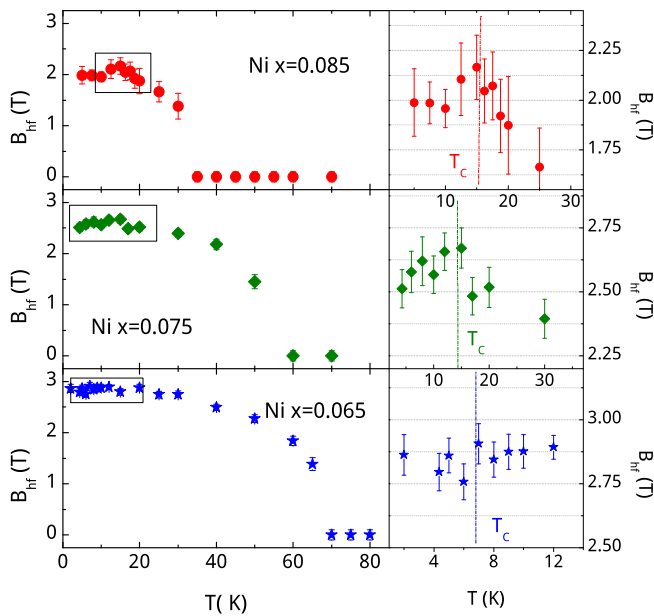}
\caption{\label{fig5} Magnetic hyperfine field $B_{hf}$ for BaFe$_{2-x}$Ni$_x$As$_2$ single crystals as a function of temperature.  The region marked by a box on the left side is shown on an expanded scale at the right side.  $T_C$ is indicated by vertical lines. }
\end{figure}

Magnetic hyperfine fields $B_{hf}$ obtained by $^{57}$Fe M\"ossbauer spectroscopy fits for BaFe$_{2-x}$Ni$_x$As$_2$ single crystal mosaics are shown in figure \ref{fig5}.  Making the simple approximation that $B_{hf}(0)$ is proportional to the Fe magnetic moment \cite{dubiel} we observe a clear systematic reduction of the magnetic moment compared to the parent compound caused by doping.  The temperature dependence of $B_{hf}$ shows a change at $T_C$ as follows: a decrease of $B_{hf}$ clearly is observed for x = 0.075 and x= 0.085 Ni doping, while for x= 0.065 there is no clear variation of $B_{hf}$ near $T_C$, only a saturation of $B_{hf}$ at low temperatures is observed.  As already mentioned above, a similar behavior had been seen in the same Ni doped single crystals by neutron scattering \cite{pratt,christianson,mwang1,mwang2}: a decrease in the Bragg peak intensity and a spin resonance at $T_C$ has been reported.  The reduction observed in our $B_{hf}$ data seems to be directly related to the difference between $T_N$ and $T_C$, the smaller the difference the larger the decrease of $B_{hf}$.  A possible reason is that the competition between superconductivity and magnetic order becomes stronger when the energy scales for the magnetic and superconducting transition are of the same order.  Another explanation may be as follows: if the magnetic ordering temperature $T_N$ is much higher than $T_C$ a possible phase separation between a magnetic and a superconducting phase will result in a much larger length scale of phase separation than for the case that the difference between $T_N$ and $T_C$ is small.  The length scale of a possible phase separation, therefore, will decrease with increasing Ni doping x, i.e. going from x = 0.065 to x= 0.085.  A change in the hf data of the magnetic phase at $T_C$ only can be observed if the length scale of such a possible phase separation is smaller than the superconducting coherence length $\zeta_{SC}$ (2 nm\cite{putti}).  We, therefore, can conclude, that if there is a phase separation, its length scale will be smaller than $\zeta_{SC}$ for x = 0.085 and 0.075, while it will be larger than $\zeta_{SC}$ for x = 0.065.

It is well known that, in the iron pnictide systems, the Fermi surface for the parent compounds is composed by two concentric electron pockets centered at ($\pi$,$\pi$) and two concentric hole pockets centered at (0,0) in the Brillouin zone, and nesting between electron and hole pockets occurs, giving rise to SDW order.  Superconductivity occurs when nesting is weak, caused by distortion of the Fermi surface, that is, through doping or external pressure \cite{chan}.  When the system is doped with electrons, the electron pockets expands and the hole pockets contracts, and the inverse happens when the system is hole doped.  This Fermi surface distortion reduces nesting and gives possibility to appearance of Cooper interactions between electrons \cite{fawang}.  In the light of this information, we can think that for the case we have magnetic SDW ordering and superconductivity in the same sample, we can have phase separation with one of the phases still showing Fermi surface nesting and thus magnetic ordering, while for the other phase the Fermi surface nesting can be broken giving rise to superconductivity.  On the other hand, we could also expect an overlap between magnetism and superconductivity, that means we would have at the same time conditions to have Fermi surface nesting for weakened magnetism and favorable conditions to have Cooper pairing.  %, due to that weak coupling between electron and hole pockets.

The data presented in figures \ref{fig3}, \ref{fig4} and \ref{fig5} indicate that there is a connection between magnetic ordering and superconductivity: a decrease of the magnetic hyperfine field $B_{hf}$ below $T_C$ clearly can be observed.  Such a decrease only can be seen if we have either coexistence between magnetism and superconductivity or a phase separation on a length scale smaller than the superconducting coherence length $\zeta_{SC}$.  Since $\zeta_{SC}\sim 2$ nm\cite{putti} which is of the order of the unit cell, it obviously does not make sense to talk about a real ``phase separation''.  Coexistence between magnetism and superconductivity, therefore, seems to be established from our M\"ossbauer studies.  Neutron scattering studies on the same samples \cite{pratt,christianson,mwang1,mwang2} show a decrease in the Bragg peak intensity accompained by a resonance below $T_C$.  This means there is a decrease in the Fe magnetic moment or in the magnetic volume fraction accompained by a change in the Fe magnetic moment dynamics.  Taking our M\"ossbauer results which clearly do not see a change in the magnetic volume fraction below $T_C$, but a decrease in the magnetic hyperfine field, we can conclude that there is a decrease in the Fe magnetic moment below $T_C$.  It is argued that such a decrease is caused by a spectral weigth transfer when entering the superconducting state.  It can be explained by assuming $s^{+-}$ pairing symmetry, where reentrance of the nonmagnetic phase occurs below $T_C$ and thus reducing the Fe magnetic moment\cite{rafael}.

Concluding we can say that our $^{57}$Fe M\"ossbauer studies give evidence for the coexistence of magnetism and superconductivity in single crystals of Ba$_{0.78}$K$_{0.22}$Fe$_2$As$_2$ and BaFe$_{2-x}$Ni$_x$As$_2$ for x= 0.075 and 0.085 which is expressed in a reduction of the Fe magnetic moment below $T_C$.  This clearly shows the advantages of using M\"ossbauer spectroscopy compared to other techniques, e. g. neutron scattering or $\mu$SR: (i) $^{57}$Fe M\"ossbauer spectroscopy as a local technique can distinguish between the change of the local magnetic moment and that of the magnetic volume fraction, respectively; this, however, is not possible by neutron scattering; (ii) $^{57}$Fe M\"ossbauer spectroscopy measures the magnetic moment directly at the probe (Fe) atom, while in $\mu$SR the site were the muon is coming at rest may be one or several sites away from the Fe position.  We obtain additional information on the reduction of the Fe magnetic moment when entering the superconducting state: this reduction seems to depend on the magntitude of the difference between magnetic ordering temperature $T_N$ and superconducting transition $T_C$.  We offer two different explanations for this observation (see above), but additional experimental data definitely are needed to decide which is the correct one.

%\begin{figure}
%\includegraphics[width=9cm]{figure6n.eps}
%\caption{\label{fig6} Paramagnetic volume fractions obtained from $^{57}$Fe M\"ossbauer spectra for BaFe$_{1.925}$Ni$_{0.075}$As$_2$ single crystal mosaics.  Inset figure shows the paramagnetic volume fraction in function of doping for BaFe$_{2-x}$Ni$_x$As$_2$, and the shaded regions correspond to doping where no coexistence of magnetism and superconductivity had been observed. }
%\end{figure}

%Putting all these things together, what we have is that Fe is sensing magnetic ordering below $T_N$ and is also sensing variation in its dynamics at and below $T_C$, either for electron and hole doping.  We are also seeing regions where no magnetism is present.

{\bf Acknowledgement:\/}  This work has been supported by the CNPq (under CIAM collaboration with NSF) and FAPERJ agencies at CBPF in Rio, and by US NSF under the Materials World Network (MWN: DMR-0806846) and the Partnership for International Research and education (PIRE: OISE-0968226) programs at Columbia.  H. Micklitz acknowledge visitor fellowship of CAPES to work at CBPF.  We acknowledge Mariella Alzamora for her help during the preparation of the single crystal mosaics.  C. Arg\"uello acknowledge APS Fellowship for his visit to Rio.\\

\vfill \eject
\end{document}